\documentclass[fleqn,10pt]{wlscirep}
\usepackage[utf8]{inputenc}
\usepackage[T1]{fontenc}
\usepackage{booktabs}
\title{Common-Mode Control and Confinement Inversion of Electrostatically Defined Quantum Dots in a Commercial CMOS Process}
\author[1,*,+]{Andrii Sokolov}
\author[1,2,+]{Xutong Wu}
\author[1,2,+]{Conor Power}
\author[3]{Mike Asker}
\author[1]{Panagiotis Giounanlis}
\author[1]{Ioanna Kriekouki}
\author[4]{P\`{e}ter Hanos-Puskai}
\author[1]{Conor McGeough}
\author[3]{Imran Bashir}
\author[1]{David Redmond}
\author[3]{Dirk Leipold}
\author[1,2]{Bogdan Staszewski}
\author[1,2]{Elena Blokhina}

\affil[1]{Equal1 Laboratories, Dublin, D04 V2N9, Ireland}
\affil[2]{Centre for Quantum Engineering, Science, and Technology and School of Electrical and Electronic Engineering, University College Dublin, Dublin, D04 V1W8, Ireland}
\affil[3]{Equal1 Laboratories, San Carlos, CA 94070, California, USA}
\affil[4]{Equal1 Laboratories Romania SRL, 300124, Timișoara, Romania}

\affil[*]{andrii.sokolov@equal1.com}

\affil[+]{these authors contributed equally to this work}

\begin{abstract}
  Confining electrons or holes in quantum dots formed in the channel of industry-standard fully depleted silicon-on-insulator CMOS structures is a promising approach to scalable qubit architectures. In this article, we present our results on a calibrated model of a commercial nanostructure using the simulation tool Quantum TCAD, along with our experimental verification of all model predictions. We demonstrate here that quantum dots can be formed in the device channel by applying a combination of a common-mode voltage to the source and drain and a back gate voltage. Moreover, in this approach, the amount of quantum dots can be controlled and modified. Also, we report our results on an effective detuning of the energy levels in the quantum dots by varying the barrier gate voltages. Given the need and importance of scaling to larger numbers of qubits, we demonstrate here the feasibility of simulating and improving the design of quantum dot devices before their fabrication based on a commercial process.

\end{abstract}

\begin{document}

\flushbottom
\maketitle

\thispagestyle{empty}

\section*{\label{sec:introduction} Introduction}

Quantum dots formed through electrostatic means and material interfaces are the basis of one of the approaches to quantum computing hardware based on semiconductor materials\cite{chatterjee_semiconductor_2021}. Various degrees of freedom can be used to encode quantum information in these quantum dots, such as the charge or spin, in order to form a set of qubits. More exotic multi-particle spin states acting as qubits have also been demonstrated in semiconductor quantum dots, such as singlet/triplet, flip-flop, and hybrid states~\cite{burkard_semiconductor_2023}. Fundamental to all of these implementations is the precise electrostatic control over the quantum dots and the barriers between them. A typical semiconductor structure hosting two quantum dots separated by barrier gates\cite{vahapoglu_coherent_2022} has the following components: two electron reservoirs for the injection of charge carriers from either end of the quantum dot array, three barrier gates to control tunnelling between adjacent quantum dots, and two plunger electrodes between the gates to control the electrochemical potential of the quantum dots.  

In this paper, we focus on the modelling and experimental characterisation of a quantum dot array (QDA) in a commercial 22~nm Fully Depleted Silicon-On-Insulator (FDSOI) process from GlobalFoundries~\cite{ong_22nm_2019}. As opposed to conventional semiconductor quantum dot arrays\cite{seidler_conveyor-mode_2022, burkard_semiconductor_2023}, our structure does not have any plunger electrodes between the gates. We demonstrate full electrical control over the location of the quantum dots, either underneath or between the gate electrodes, through a common-mode voltage applied to the source and drain terminals of the QDA and the barrier gate voltages. Confinement in other semiconductor QDAs is typically done below gate or plunger terminals\cite{west_gate-based_2019,ciriano-tejel_spin_2021, piot_single_2022}, however here we demonstrate confinement between barrier gate terminals without a dedicated plunger terminal. An effective plunger is then realised through variations in the common-mode voltage and the barrier gate voltages depending on the region of operation. %
Biasing through the common-mode voltage allows for moderate biasing voltages, given the top gates and source/drain regions in this device are not frozen out at 1~K (see Section.~\ref{sec:simulation} for further discussion). The quantum well formation for a given biasing condition is accurately predicted in simulation and is confirmed with a range of experimental demonstrations. Thus, we also demonstrate the feasibility of simulating quantum transport properties of commercial semiconductor devices prior to fabrication. This is a key requirement for the design of future scalable quantum computing architectures using commercial semiconductor technologies\cite{costa_advances_2023, vinet_path_2021}.

The remainder of the paper is structured as follows: Section~\ref{sec:system_overview} contains an overview of the device, including limitations of the commercial photo-lithography process. In Section~\ref{sec:simulation}, we describe the semiconductor and quantum mechanical model used to predict the behaviour of the quantum dots in the device. Sections~\ref{sec:flatband_sim} and \ref{sec:sim_detuning} describe the simulation results using our model. Finally, Section~\ref{sec:measurement_setup} and Section~\ref{sec:measurement_results} are dedicated to the measurement setup and results that support the predicted operation of the device, such as charge stability diagrams, Coulomb blockade effects at 1~K, and bias triangle pair formation.

\section{\label{sec:system_overview} System Overview
}
\begin{figure}[htb]
\centering
\includegraphics[width=0.48\linewidth]{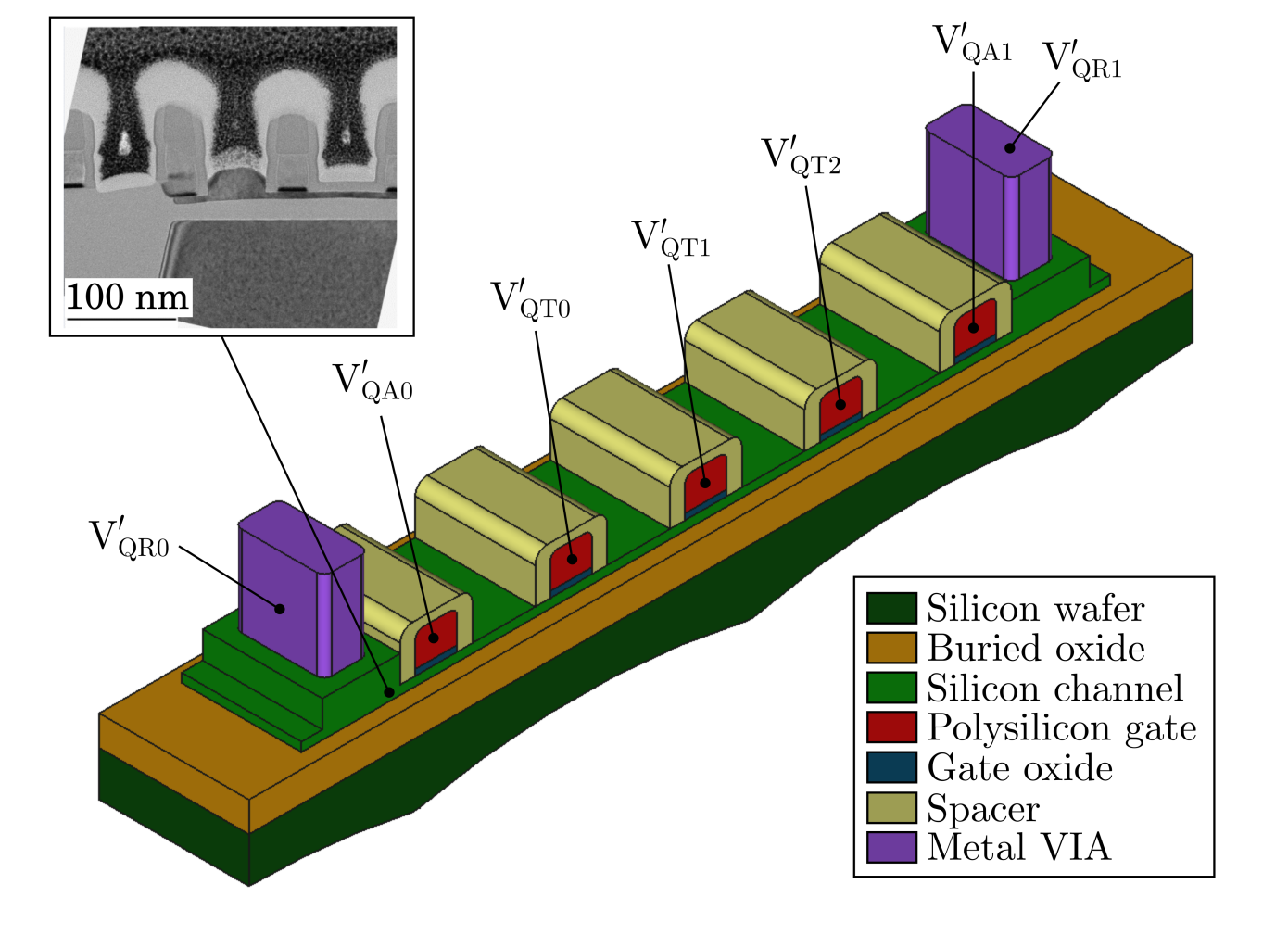}
\caption{\label{fig:sys_figure} 3D view of the five-gate quantum dot array with raised source and drain. The scanning electron microscope (SEM) image of the device shows dummy polysilicon gates that are not included in the 3D view. A backgate terminal is also available in this process but is not visible in this diagram. It connects through a metal VIA to the silicon wafer below the buried oxide.}
\end{figure}

The device investigated in this study is fabricated using GlobalFoundries' 22~nm Fully Depleted Silicon-On-Insulator (FD-SOI) technology. As illustrated in Fig.\ref{fig:sys_figure}, the device consists of a raised source and drain, and five electrostatic gates and it serves as an initial test structure for future highly scalable QDAs based on industry-standard FD-SOI processes. We will explain the challenges encountered throughout the paper relating to the electrical control of the QDA structure.

\begin{table}[h!]
\centering
\begin{tabular}{@{}lll@{}}
\toprule
Voltage & Description & Equation (if applicable) \\
\midrule
$V'_\text{QR0}$ & Source potential &  \\
$V'_\text{QR1}$ & Drain potential & \\
$V'_\text{QA0}$ & QA0 potential &  \\
$V'_\text{QA1}$ & QA1 potential &  \\
$V'_\text{QT0}$ & QT0 potential &  \\
$V'_\text{QT1}$ & QT1 potential &  \\
$V'_\text{QT2}$ & QT2 potential &  \\
\midrule
$V_\text{CM}$ & Common-mode voltage & $\left( V'_\text{QR0} + V'_\text{QR1} \right)/2$ \\
$V_\text{DS}$ & Drain-to-source voltage & $V'_\text{QR1} - V'_\text{QR0}$ \\
$V_\text{QA0}$ & QA0 potential w.r.t source & $V'_\text{QA0} - V'_\text{QR0}$ \\
$V_\text{QA1}$ & QA1 potential w.r.t source & $V'_\text{QA1} - V'_\text{QR0}$ \\
$V_\text{QT0}$ & QT0 potential w.r.t source & $V'_\text{QT0} - V'_\text{QR0}$ \\
$V_\text{QT1}$ & QT1 potential w.r.t source & $V'_\text{QT1} - V'_\text{QR0}$ \\
$V_\text{QT2}$ & QT2 potential w.r.t source & $V'_\text{QT2} - V'_\text{QR0}$ \\
\bottomrule
\end{tabular}
\caption{Terminal voltages and potentials}\label{tab1}%
\end{table}

As can be seen in Fig.\ref{fig:sys_figure}, unlike other QDAs with quantum dots controlled by plunger gates\cite{burkard_semiconductor_2023}, this device admits indirect electrical control over quantum dots forming between the polysilicon barrier gates. The fabrication of the device does not require modification of the standard photolithography process\,\cite{wu_evolution_2020, hou_optical_2021, park_patterning_2023}. We emphasise that the most pertinent feature of the standard processing that we must adhere to is standard sizing and pitch, which places strict constraints on the design of quantum dot devices in photolithography-fabricated devices. In our structure, the distance between neighbouring gates is approximately $1.5$ to $2$ times the size of the resulting quantum dots. The gates QA0 and QA1 act as spacers between the heavily doped source and drain and the quantum dots. Therefore, the electrical control scheme becomes more complicated compared to traditional structures with plungers as each gate affects both tunnel barriers and quantum dot energy levels. Later in the paper, we describe simulations and measurement results that demonstrate the biasing and detuning of the device without plunger gates.

When designing an experiment to control the formation of quantum wells in the channel of the device, it is necessary to keep the gate-to-source and drain-to-source voltage below 1.6 V, beyond which the sample would likely be damaged due to the breakthrough of the thin ($\approx 10$~nm) silicon nitride spacer layer. Thus, the potentials applied to the terminals are applied with respect to the source (Tab.~\ref{tab1}). All biasing conditions given with respect to the source are denoted without a prime (Tab.~\ref{tab1}). Throughout the rest of the paper, we will prove that the number of quantum dots and effective detuning can be implemented in a commercial FD-SOI device using only a common-mode voltage and barrier gate voltages through simulation and experimental verification.

\section{Simulation techniques and setup}\label{sec:simulation}

To examine if the number of quantum dots can be controlled through variations of the common-mode voltage and barrier gate voltages, we begin by building a model of our semiconductor nanodevice. However, the convergence of semiconductor equations below 70~K is highly numerically unstable due to the exponential dependence of ionised charge carriers on temperature~\cite{beckers_cryogenic_2018, gao_quantum_2013}. Therefore, simulation results obtained below this temperature typically do not coincide with an experimental characterization of semiconductor transistors. For this reason, we used the highly specialized tool kit, Quantum TCAD (QTCAD) by Nanoacademic Technologies~\cite{noauthor_qtcad_nodate}, allowing for semiconductor quantum dot modelling at deep cryogenic temperatures~\cite{beaudoin_robust_2022, kriekouki_interpretation_2022}. QTCAD allows one to solve the Poisson and self-consistent Poisson-Schrodinger equations with the assumption that classical transport is forbidden in the fully depleted ``dot region'' where quantum dots are formed. In addition, one can compute single-electron wavefunctions in these dot regions, extract lever arms with respect to device terminals, and compute sequential tunnelling current. Such a model allows for the prediction of experimental biasing conditions for various operating modes of a QDA which we discuss throughout the rest of this paper.

In our current QTCAD model, we use the following assumptions: 
\begin{itemize}
\item The gate polysilicon volumes are assumed to be conductive at cryogenic temperatures and are replaced by a set of equipotential boundaries with potentials and work functions. This approach significantly reduces the number of points to calculate without the introduction of a numerical error. This is a typical approach in TCAD modelling~\cite{maiti_introducing_2017}.
\item Both nitride and foamed spacers are considered to be perfect insulators with dielectric constants of 9 and 2.7 respectively. In reality, both of these materials are non-crystalline and therefore have very complicated band structures. However, since we are focused on the silicon channel, we neglect this complication as we are not considering the effect of charges in the insulating regions. The exact material of the foamed spacer is not known but is likely foamed $\text{SiO}_{2}$ or foamed silicon nitride based what we have found through calibration of our model against experimental data.
\item The bottom of the buried oxide region is considered a frozen boundary which is applicable when the thermal energy $k_\text{B}T$ is much lower than the donor (acceptor) binding energy. This places the Fermi level in this region between the donor (acceptor) level and the conduction (valence) band edges~\cite{pierret_advanced_2003}.
\item Source and drain metal contacts are considered ohmic boundaries. The source and drain boundary conditions are set by locally shifting the Fermi energy level by $e V_\text{source}$ and $e V_\text{drain}$ respectively, where $e$ is the electron charge.
\item Mechanical stress in the silicon channel is neglected in this simulation. This can introduce a noticeable offset between the biasing voltages predicted by simulation and those used in the experiment\cite{maiti_introducing_2017}. This will be considered in future work.
\end{itemize}

The calibration of the QTCAD model is central to its use in predicting biasing conditions. Without accurate calibration, we can learn very little from our model. The unknown parameters that we derive from the experiment are noted as follows: the doping concentration ($n_\mathrm{sd}$) in the raised source and drain (in reality it is a function of space $n_\mathrm{sd}=n_\mathrm{sd}(x,y,z)$ but is assumed constant over the drain/source volumes here), the gate work function $E_\mathrm{Wg}$, the back-gate work function $E_\mathrm{Wbg}$, and the doping concentration $n_\mathrm{bg}$ of the frozen back-gate region. Initial information for building our model was taken from publicly available process information, actual commercial processes may differ somewhat\cite{maiti_introducing_2017}. In our initial calibration, we used experimentally obtained $i_\text{DS}(V_\text{CM})$ curves and qualitatively compared them with conduction band configurations from simulation. This allows us to reduce the possible range of many parameters in the semiconductor simulation. The gate polysilicon is assumed to be p-doped\cite{maiti_introducing_2017}.
The simulation of a Gaussian doping distribution in the source and drain introduces an unreasonably high-density Finite Element Method (FEM) mesh. Therefore, uniform doping was assumed in all volumes of the source and drain domains. It is unclear what an equivalent doping concentration is in this approach, so we used a wide-range scan of the doping concentrations to match experimental results (See supplementary materials). We also used the standard frozen boundary condition on the back gate\cite{noauthor_qtcad_nodate}. 

A classification of the conduction bands (Fig.~\ref{fig:sim_figure} and
Tab.~\ref{tab:device_states_classification}) allows further calibration of the
QTCAD model. A finer calibration was then done using an experimentally obtained
Coulomb blockade in a minimum size 22~nm FDSOI transistor. We used biasing
conditions that correspond to a single visible Coulomb diamond and the measured
lever arm to fine-tune the simulator parameters (See supplementary materials for
more details). 

\section{Flat Band Simulation and Well Location Characterisation}
\label{sec:flatband_sim}

\begin{figure}[h]
\centering
\includegraphics[width=0.9\linewidth]{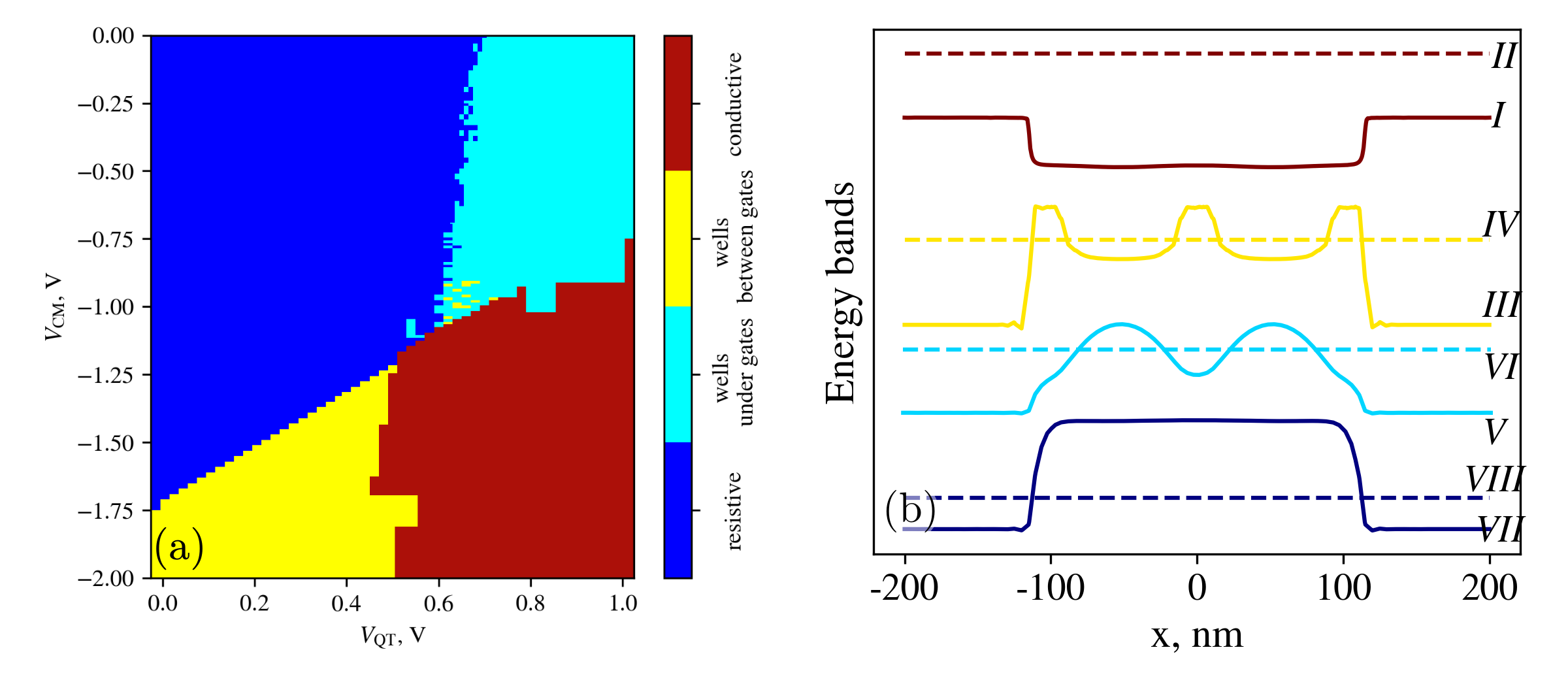}
\caption{\label{fig:sim_figure} 
(a) Qualitative analysis of the flat band simulation. Each coloured pixel of the diagram corresponds to the broad shape of the conduction band in the channel of the device as per the device states noted in Tab.\,\ref{tab:device_states}. 
(b) Examples of variation of the band diagrams under different bias conditions. I, II --- conduction band and Fermi energy level of the device in conductive regime; III, IV --- quantum wells are formed between barrier gates; V, VI --- one quantum dot is formed under QT1 (under barrier gate); VII, VIII --- a global barrier higher than Fermi level forms, thus the device is non-conductive. The effect of the voltages $V_{\text{QA0}}$ and $V_{\text{QA1}}$ are not shown in Fig.~\ref{fig:sim_figure}. These conduction band slices are only examples of specific bias conditions.}
\end{figure}

The first stage of the device characterization was a flat band simulation and experiment. This is a sweep of $V_\text{CM}$ with all equal gate-to-source barrier voltages, $V_\text{QT} = V_\text{QT0} = V_\text{QT1} = V_\text{QT2}$. The experimental results are reported in Sec.~\ref{sec:measurement_results}. In the simulation, each biasing point was classified according to Tab.~\ref{tab:device_states}/Tab.~\ref{tab:device_states_classification} and the results are shown in Fig.~\ref{fig:sim_figure}~(a), with a unique colour for each device state. The conduction band edge has a complex 3D structure, with its cross-sections varying from the top of the channel close to the gate/channel interface down to the bottom of the channel, close to the buried oxide interface. Therefore, a quantitative classification into each of the states in Tab.~\ref{tab:device_states} is described in Tab.~\ref{tab:device_states_classification}.

\begin{table*}[htb]
\centering
\begin{tabular}{@{}p{2.4cm}p{6cm}p{2.4cm}p{2.4cm}@{}}
\toprule
Device state & Description &Conduction band in Fig.~\ref{fig:sim_figure}~(a) & Fermi level in Fig.~\ref{fig:sim_figure}~(a) \\
\midrule
Conductive & All barriers are lower than the Fermi energy level, the drain and source voltages are high.  &I & II \\
Wells between gates & Potential wells form between the barrier gates. The device allows for a tunnelling current. & III & IV \\
Wells under gates & The potential well forms under the barrier gates. The device allows for a tunnelling current. &  V & VI\\
Non-Conductive & There is a wide potential barrier between source and drain. The current between the source and drain is very low &  VII & VIII \\
\bottomrule
\end{tabular}
\caption{Qualitative classification of conduction bands.}\label{tab:device_states}%
\end{table*}

\begin{table*}[htb]
    \centering
    \begin{tabular}{@{}p{4cm}p{12cm}@{}}
    \toprule
    Device State & Classification Test \\
    \midrule
    Conductive & A linecut is taken along the top of the channel of the device, near the barrier gate/channel interface. If the Fermi level is larger than the conduction band edge linecut the state is classified as conductive, otherwise we move to the remaining states to classify. \\
    Wells between gates/Wells under gates &  The most probable location of the single-electron groundstate found by the Schrodinger solver is first computed. If this lies between the barrier gates, then the state is classified as such. Otherwise, if the most probable location lies underneath a barrier gate then the state is classified as wells under gates.\\
    Non-Conductive & The energy of the single-electron groundstate is compared to the Fermi level. If this energy is larger than the Fermi level, then the state of the device is classified as non-conductive.\\
    \bottomrule
    \end{tabular}
    \caption{Tests used for classification of conduction bands.}\label{tab:device_states_classification}%
\end{table*}

Figures~\ref{fig:sim_figure}~(a),\,(b) show the four distinct modes of operation of the device as described in Tab.~\ref{tab:device_states}. A varying number of quantum dots form in the channel of the device in two of the states, which can be controlled by $V_{\text{CM}}$ and $V_{\text{QT}}$:

\begin{itemize}
\item In the case of a quantum dot forming under QT1 (curves V and VI in Fig.~\ref{fig:sim_figure}~(b)), the voltage applied to QT1 changes the energy levels of an electron trapped in this quantum dot. The ratio between the variation in the energy level in a quantum dot and the variation in the gate voltage is known as the lever arm. The potential barrier height separating the central quantum dot from the source and drain regions is controlled by the common-mode voltage and the two side barrier gate voltages, $V_{\text{QT0}}$ and $V_{\text{QT2}}$.
\item In the case when quantum dots form between barrier gates (curves III and IV in Fig.~\ref{fig:sim_figure}~(b)), the voltage applied to the barrier gate terminals controls the corresponding potential barriers between the quantum dots and the energies of the quantum dots to a lesser extent. The common-mode voltage primarily controls the energy of both quantum dots as an effective global plunger. However, for independent control of the energy levels in each of the two dots, we describe effective detuning later in this paper using the barrier gate voltages.
\end{itemize}

Given the operating modes noted above, the energy levels present in the quantum dots change with variations in both $V_\mathrm{QT}$ and $V_\mathrm{CM}$, which is verified experimentally. The choice of biasing conditions that prepare the device in one of the four states noted in Tab.\,\ref{tab:device_states} is dependent on the intended use. The creation of a single quantum dot under QT1 results in a quantum dot that is well isolated from the source and drain regions due to the large potential barriers. When the device is operated in the regime with quantum dots appearing between QT0/QT1 and QT1/QT2, we have two quantum dots with a controllable inter-dot potential barrier (QT1), and two more potential barriers separating the dots from the source and drain (QT0, QT2). In this latter case, the barriers are not as wide as in the case of the single quantum dot and so the dot is more strongly coupled to the source and drain. Later in this paper, we utilize the fact that the barrier gate voltages affect the energy of nearby dots to control the energy levels in individual quantum dots.

\section{Effective Detuning of Quantum Dots in Simulation}\label{sec:sim_detuning}

\begin{figure}[h!]
    \centering
    \includegraphics[width=0.75\textwidth]{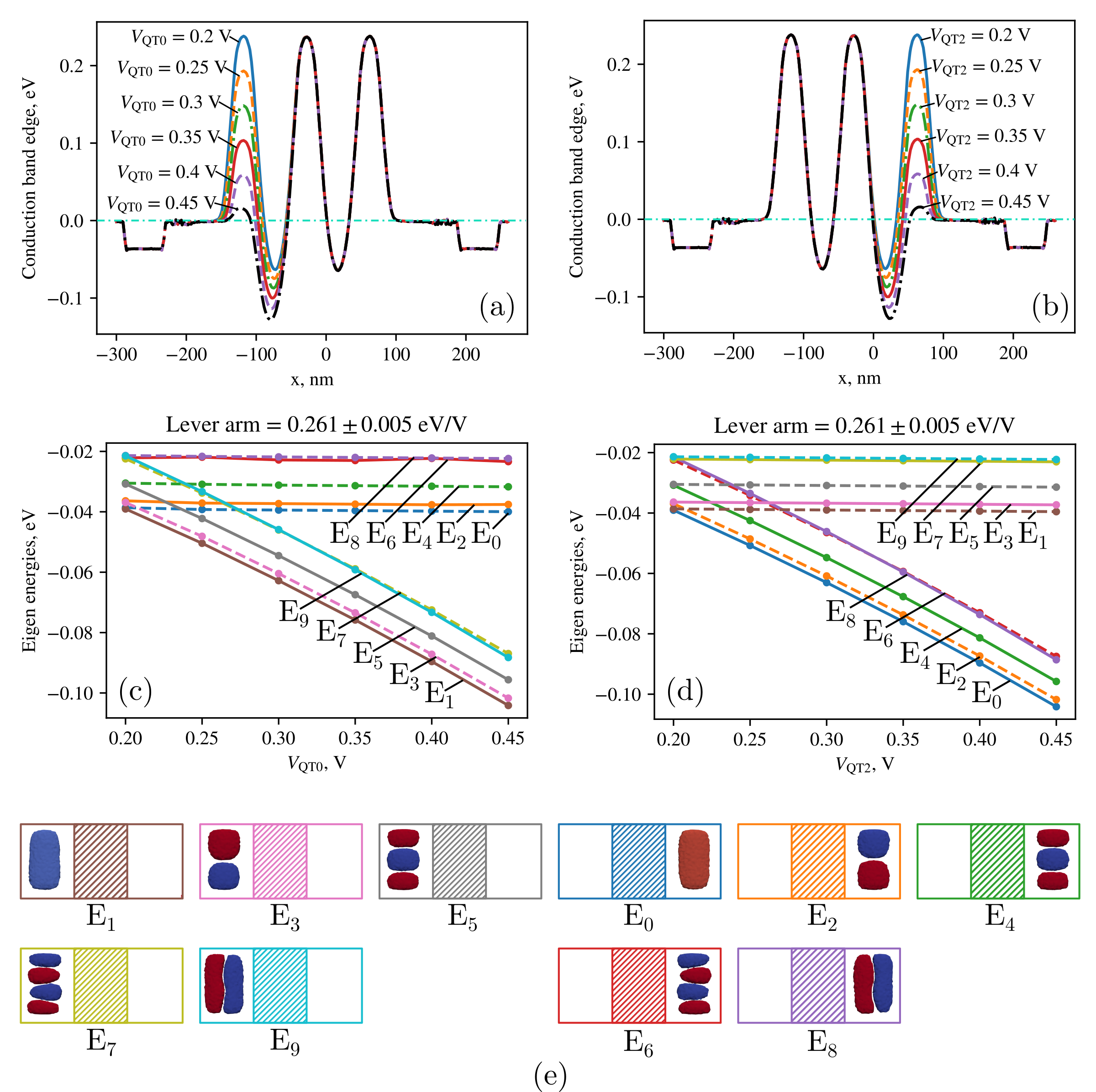}
    \caption{Effective detuning of quantum dots forming between QT0/QT1 and QT1/QT2 by sweeping the voltages $V_{\text{QT0}}$ and $V_{\text{QT2}}$ respectively. (a),(b) show the conduction band and its response to sweeps in $V_{\text{QT0}}$ and $V_{\text{QT2}}$ respectively. The horizontal line shown in both conduction band plots at $0~\text{eV}$ is the Fermi level. (c),(d) define leverarms capturing the linear response of states bound in the left and right quantum dots to sweeps in $V_{\text{QT0}}$ and $V_{\text{QT2}}$ respectively. Notice that $V_{\text{QT0}}$ has very little effect on states bound in the quantum dot that is not directly adjacent to it. The same is also true for $V_{\text{QT2}}$ and the left quantum dot. (e) shows the first 10 single-electron orbital states and their location in the double quantum dot system.}
    \label{fig:QT_sweeps}
\end{figure}

Given that we can bias our device as described in Tab.\,\ref{tab:device_states}, we now describe the process of detuning our device in the double quantum dot state by sweeping the barrier gate voltages, $V_{\text{QT0}}$ and $V_{\text{QT2}}$, whilst keeping all other voltages fixed. This allows us to demonstrate effective plungers that control the energy levels of the two quantum dots independently, as illustrated in Fig.~\ref{fig:QT_sweeps}.

Figure~\ref{fig:QT_sweeps}(a) shows the variation in the conduction with variations in the voltage $V_{\text{QT0}}$. The dips in the conduction band to the left and right of the central three-barrier structure are due to $V_{\text{QA0}}$ and $V_{\text{QA1}}$. The voltage $V_{\text{QT0}}$ is swept from $0.2~\text{V}$ to $0.45~\text{V}$. Note the variation in the barrier height and the well depth. The change in the depth of the well is approximately linear, as shown in Fig.~\ref{fig:QT_sweeps}(c), where the change in single-electron orbital energies in the left quantum dot is linear. Therefore, we can define a leverarm to capture this linear response of the left-bound states with respect to variations in $V_{\text{QT0}}$ of approximately $0.261~\text{eV/V}$.

The same simulation is then also performed by sweeping $V_{\text{QT2}}$ over the same range. The conduction band response and state energy variation are shown in Fig.~\ref{fig:QT_sweeps}(b),(d). Note again that the states bound in the left quantum dot are essentially unaffected by variations in $V_{\text{QT2}}$ while states bound in the right quantum dot respond with a leverarm relation of approximately $0.261~\text{eV/V}$. Therefore, simulation suggests that the barrier gate voltages $V_{\text{QT0}}$ and $V_{\text{QT2}}$ can act as plungers, directly controlling the energy of the bound states in the left and right quantum dots respectively.

We can also estimate the size of the wavefunctions and see how they compare to the pitch of the barrier gates in this device. For a quantum dot forming between two barrier gates separated by a pitch of 90nm the ground state wavefunction has the following extent in the X,Y and Z directions (allowing for 3 standard deviations around its central point): [15nm 47nm 3nm]. Even though the ground state wavefunction size is not that comparable to the pitch, we do show clear single and double dot control in Sec.~\ref{sec:measurement_results} through higher energy states which would typically be larger in size and more comparable to the pitch.

We also note that there is excellent agreement between the predicted leverarm in QTCAD simulation ($\approx 0.261~\text{eV/V}$) and that measured in experiment ($\approx 0.2701~\text{eV/V}$). See the supplementary materials for Coulomb diamond measurement results and the method used to extract the leverarm.

\section{Measurement Setup\label{sec:measurement_setup}}

The test chip is placed in a cryo-cooler which is capable of reaching $40$~K, $4$~K, and $1$~K in three stages. The Quantum Machines OPX+\cite{noauthor_quantum_nodate} is used for supplying a radio-frequency (RF) reflectometry input signal from the RF carrier channel and for digitizing the reflected signal on the digitizer channel. A QDevil QDAC-II is triggered by the OPX+, and the DC voltages are loaded to the QDAC-II in advance and swept by trigger signals from the OPX+. This allows for synchronization between DC voltage sweeps and RF-reflectometry readout. Further details of the experimental setup are shown in the supplementary material. Reflectometry is performed on the source of the device which is tunnel coupled to the quantum dots forming in the channel. In this setup, the source acts as a lead for dispersive charge sensing. This is tunnel coupled to the quantum dots in the channel which act as a two-well single electron box (SEB) given that the drain lead is effectively disconnected. In this way, RF-reflectometry can then be used to detect inter-dot transitions\cite{vigneau_probing_2023}.

A flat band experiment, as demonstrated in simulation in Sec.~\ref{sec:flatband_sim}, is also measured experimentally. We then investigate the various biasing regions as noted in Fig.\,\ref{fig:sim_figure}~(a). To differentiate between single and double quantum dot confinement in the channel of our device, we first bias our device in a double quantum dot configuration. Then, by increasing $V_{\text{QT1}}$, the double quantum dot becomes a single quantum dot, and this is evidenced by the change in the measured charge stability diagram\cite{van_der_wiel_electron_2002}. We also measure bias triangle pairs to further confirm the presence of double quantum dots in the channel of a device similar to that described in Fig.~\ref{fig:sys_figure}.

\section{Measurement Results\label{sec:measurement_results}}

\subsection{Flat Band}

\begin{figure}[h]
    \centering
    \includegraphics[width=0.72\textwidth]{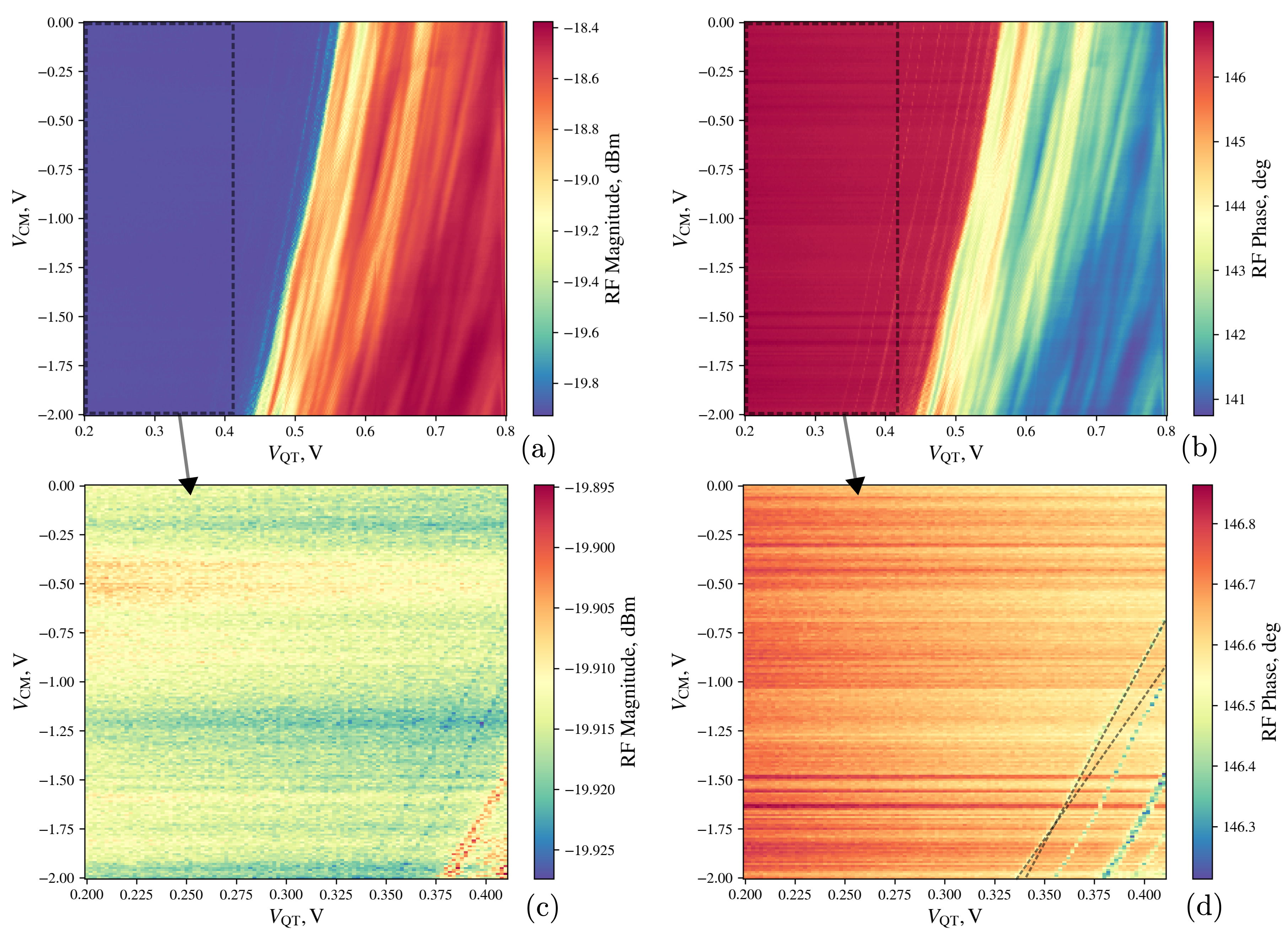}
    \caption{Flat band measurement results showing both magnitude and phase response. The characteristic slope between the conducting and non-conducting regions is clearly visible in the RF magnitude (a) and RF phase measurements (b). Zoomed-in flat band measurement results are shown in (c), and (d) to highlight Coulomb blockade transitions moving towards the fully conducting region.}
    \label{fig:exp_flatband}
\end{figure}

The flat band measurement results (Fig.\,\ref{fig:exp_flatband}) show a partitioning that is very similar to that predicted in simulations, with a characteristic slope separating the conducting and non-conducting regions. A number of Coulomb blockade transitions can be seen moving between the conducting and the non-conducting regions, shown as separated transition lines at approximately $V_{\text{QT}} = 0.4~\text{V}$. Similar flat band experiments, demonstrating Coulomb Blockade transitions, have been demonstrated in the literature~\cite{yamada_fabrication_2014}. 

A zoomed-in version of the same flat band measurement results are shown in Fig.~\ref{fig:exp_flatband}(c) and (d) to further highlight the Coulomb blockade transitions. Each transition displays a similar slope suggesting well-isolated quantum dot formation. The single and double quantum dot regimes, labelled as III and V, in Fig.~\ref{fig:sim_figure} will not necessarily admit a measurable variation in a reflectometry signal in the wide sweep explored in Fig.~\ref{fig:exp_flatband}. The main feature captured in the flatband measurement is the slope as we move into the fully conducting region, alongside some Coulomb Blockade transitions. However, as annotated in Fig.~\ref{fig:exp_flatband}(d), there is some initial evidence of a variation in the slope of the Coulomb blockade transitions in line with the variation noted in simulation (Fig.~\ref{fig:sim_figure}). A wider range of $V_{\text{CM}}$ would likely expose this variation further and will be explored in future work.

Even though this is a commercial process, disorder effects due to imperfections and charge trapping in the silicon and surrounding insulating materials are still to be expected, resulting in some of the noisy transitions seen in Fig.~\ref{fig:exp_flatband}. There is also some evidence of charge noise in the later measurements, see Fig.~\ref{fig:Triangles_0T_Loop1_Grid}. We suspect this is due to charge trapping and surface roughness at the silicon dioxide interface that the wavefunction is confined against. It is well documented in literature that this is the case in silicon processes~\cite{vinet_material_2021, martinez_variability_2022}. One advantage of the FD-SOI process is the backgate voltage, which can be used to tune the location of the quantum dot away from these interfaces. This will be explored in future work.

We expect a double quantum dot to form in the channel of the device in the bottom left corner of the flatband measurement results in Fig.~\ref{fig:exp_flatband} by comparing with the simulation results in Fig.~\ref{fig:sim_figure}(a). By the same logic, we expect a single quantum dot to form at the top boundary of the measured flatband. Next, we report the measurement results, displaying the formation and control of quantum dots in the device.

\subsection{Charge Stability Diagrams}

\begin{figure}[h]
    \centering
    \includegraphics[width=0.66\textwidth]{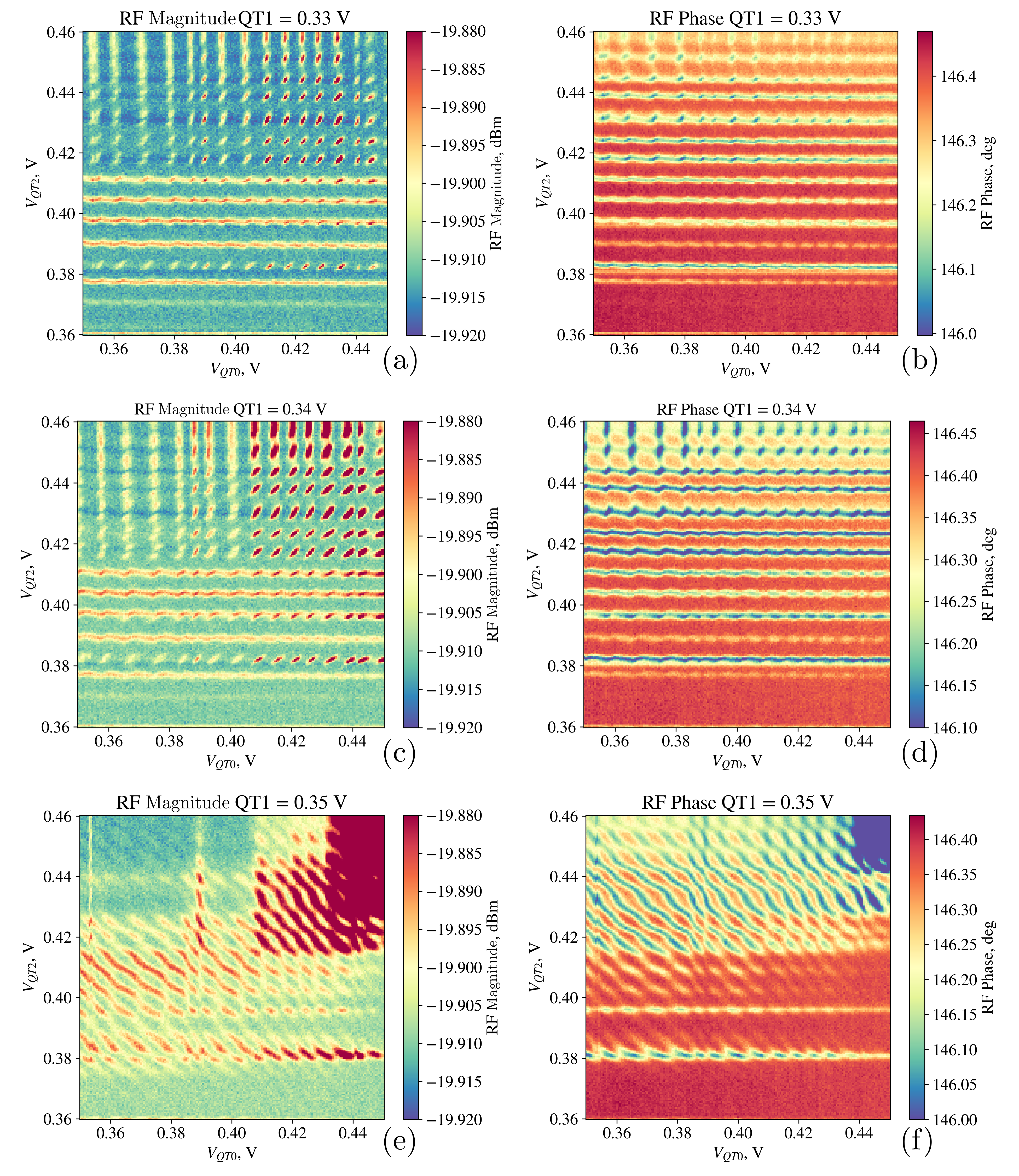}
    \caption{Measured charge stability diagrams showing the formation of a double quantum dot and a single quantum dot depending on the applied voltage of the central barrier gate $V_{\text{QT1}}$. Control over the quantum dot(s) is demonstrated by sweeping $V_{\text{QT0}}$ and $V_{\text{QT2}}$.}
    \label{fig:honeycombs}
\end{figure}

To verify the presence of single(double) quantum dot(s) forming in the channel of the FD-SOI structure, charge-stability diagrams were then measured with RF-reflectometry on the source, and are shown in Fig.~\ref{fig:honeycombs}. Charge stability diagrams show transitions between charge stable regions or equilibrium electron numbers, which varies strongly depending on the number of available quantum dots\cite{van_der_wiel_electron_2002}.

Both the magnitude and phase response of the RF reflectometry signal are shown for three values of $V_{\text{QT1}}$. As the value of $V_{\text{QT1}}$ is changed from $0.33~\text{V}$ to $0.35~\text{V}$ a clear change in the charge stability response is noted. For the lowest value, $V_{\text{QT1}} = 0.33~\text{V}$, there is a clear indication of double-quantum-dot charge transitions for $V_{\text{QT2}} > 0.42~\text{V}$ in both the magnitude (Fig.~\ref{fig:honeycombs}(a)) and phase (Fig.~\ref{fig:honeycombs}(b)) response. Given that RF reflectometry takes place on the source reservoir, which is more strongly coupled to variations due to $V_{\text{QT0}}$ than to $V_{\text{QT2}}$, we see a strong phase response connecting charge transition points in the $V_{\text{QT0}}$ direction~\cite{chorley_measuring_2012}. There is very low coupling between variations in $V_{\text{QT0}}$ and $V_{\text{QT2}}$ given the double dot stability diagram is approximately square, with charge transition lines almost horizontal and vertical.

We note that as $V_{\text{QT2}}$ increases, the coupling between the drain reservoir and the double quantum dot system increases. Therefore, as the drain to double-quantum-dot tunnel coupling increases, we expect the impedance of the system to become more sensitive to transitions between the drain reservoir and double quantum dot. This is clearly seen in Figs.~\ref{fig:honeycombs}(a),(b),(c),(d), where the lines connecting charge transition points appear completely suppressed for $V_{\text{QT2}} < 0.42~\text{V}$. For $V_{\text{QT2}} > 0.42~\text{V}$ faint lines begin to appear connecting charge transition points, quickly becoming wider as $V_{\text{QT2}}$ increases. This is more obvious for $V_{\text{QT1}} = 0.34~\text{V}$ than for $V_{\text{QT1}} = 0.33~\text{V}$ as the double quantum dot states are more strongly hybridised in the former. Therefore the overall impedance of the system, as sensed through reflectometry, shows an increased dependence on the drain reservoir coupling for $V_{\text{QT1}} = 0.34~\text{V}$.

As we increase $V_{\text{QT1}}$ to $0.34~\text{V}$ we further confirm its action as a barrier separating two quantum dots. By decreasing the potential barrier between the two quantum dots we observe a stronger coupling in the double quantum dot stability response in both the magnitude (Fig.~\ref{fig:honeycombs}(c)) and phase (Fig.~\ref{fig:honeycombs}(d)) response. Note the increased response connecting the charge transition points in both $V_{\text{QT0}}$ and $V_{\text{QT2}}$ for $V_{\text{QT2}} > 0.42~\text{V}$, especially in the phase response (Fig.~\ref{fig:honeycombs}(d)). Further increasing $V_{\text{QT1}} = 0.35~\text{V}$ shows a dramatic change in the charge stability response. We have very clearly moved from a double quantum dot charge stability response to a single quantum dot response, equally coupled to both $V_{\text{QT0}}$ and $V_{\text{QT2}}$ given the $45^{\circ}$ slope of the transition lines in Fig.~\ref{fig:honeycombs}(e),(f).

\subsection{Bias Triangle Pair Formation}

\begin{figure}[h]
    \centering
    \includegraphics[width=0.8\linewidth]{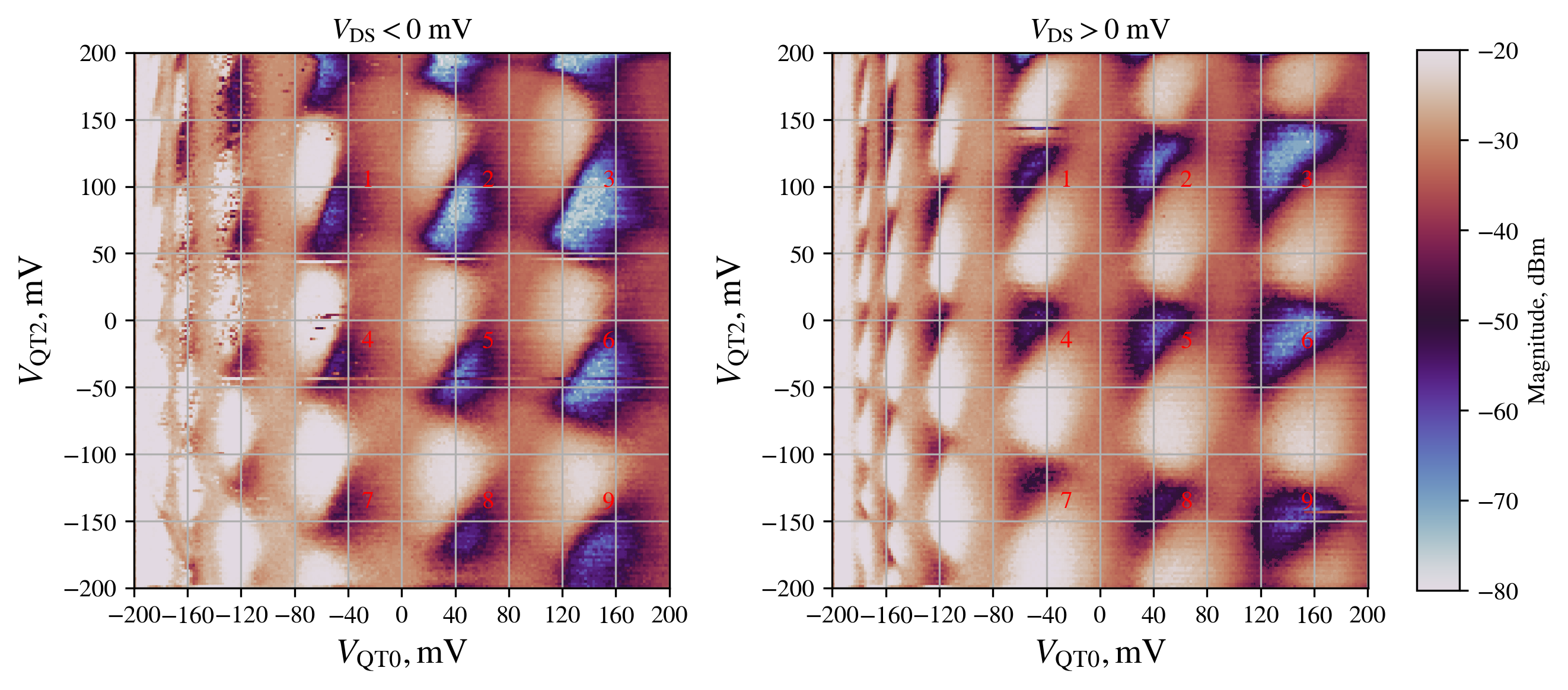}
    \caption{Charge stability sweep on voltages $V_{\text{QT0}}$ and $V_{\text{QT2}}$ with a small negative and positive $V_{\text{DS}}$. The location of the triangle pairs shifts somewhat over time so we label them $1-9$. In the negative and positive cases, $V_{\text{DS}} \approx -0.3~\text{mV}$ and $V_{\text{DS}} \approx 0.7~\text{mV}$ respectively, with the asymmetry due to an offset in the measurement setup. Improved matching and a superconducting inductor allowed for a wider range of magnitude response to be captured in this set of measurements.}
    \label{fig:Triangles_0T_Loop1_Grid}
\end{figure}

\begin{figure}[h]
    \centering
    \includegraphics[width=0.8\linewidth]{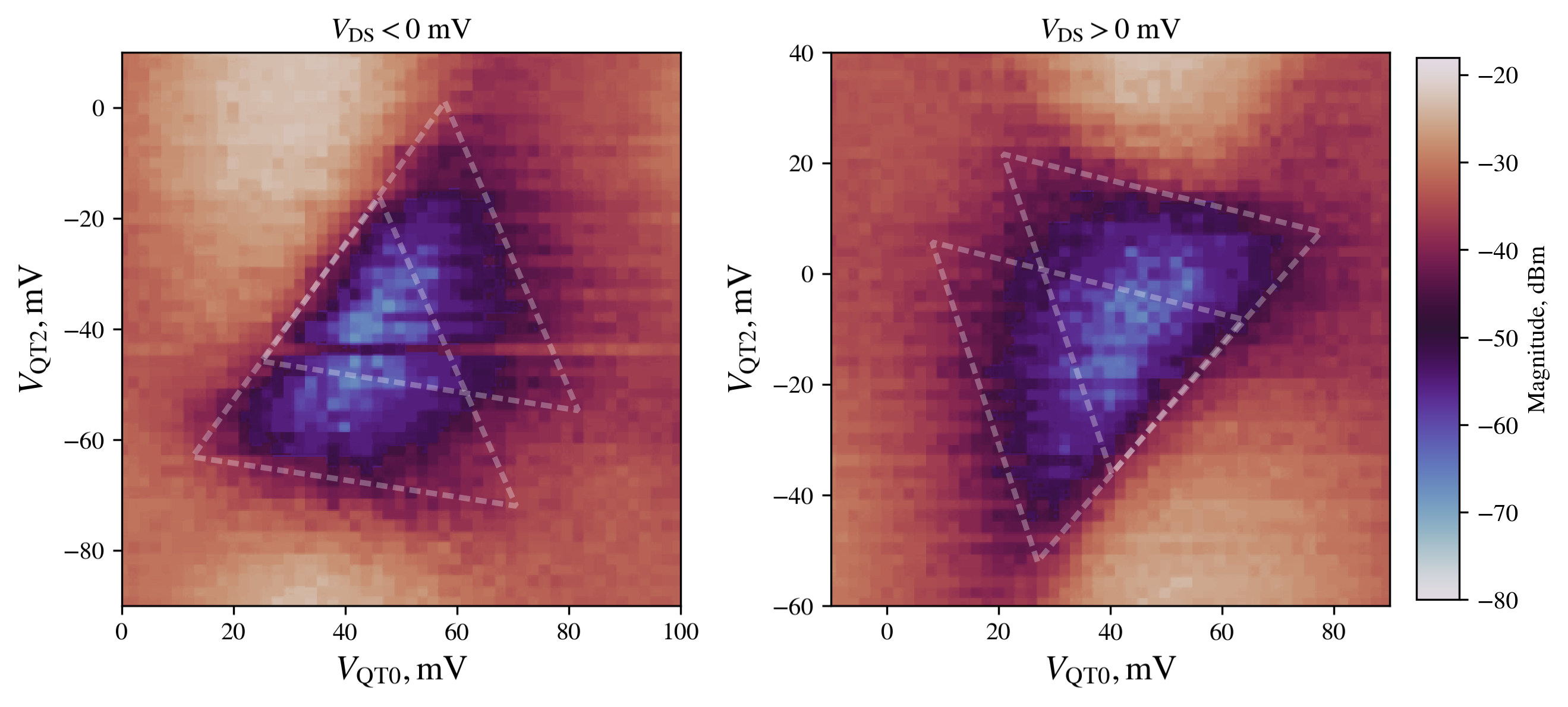}
    \caption{Zoom on bias triangles 5 from Fig.~\ref{fig:Triangles_0T_Loop1_Grid} for both positive and negative $V_{\text{DS}}$. An indication of the ideal bias triangles are overlaid on the figure.}
    \label{fig:Triangle_5_0T_Loop1_Overlay}
\end{figure}

To further verify the presence of double quantum dots in the channel of the QDA, we investigated the formation of bias triangle pairs in a device similar to that described in Fig.~\ref{fig:sys_figure} at $1$~K. These are a clear hallmark of double quantum dot transport~\cite{van_der_wiel_electron_2002}. The bias triangle pairs form at the charge transition points, such as those shown in Fig.~\ref{fig:honeycombs} with the application of $V_{\text{DS}}$. The measurement results are shown in Fig.~\ref{fig:Triangles_0T_Loop1_Grid}, with a zoomed in version of the triangles in location 5 shown in Fig.~\ref{fig:Triangle_5_0T_Loop1_Overlay}. The measurements, in this case, were taken using video-mode\cite{stehlik_fast_2015} using superconducting inductors, as longer measurements were subject to charge noise and were difficult to analyse. In this case, there is a noticeable filtering effect on the measurement data due to the nature of video-mode, causing a compression of the data for low $V_{\text{QT0}}$ in Fig.~\ref{fig:Triangles_0T_Loop1_Grid}. 

The bias triangle pairs appear for both a small negative and positive $V_{\text{DS}}$, showing a reversal in their direction as is expected\cite{van_der_wiel_electron_2002}. The formation of well-defined double quantum dots in this device, similar to that in Fig.~\ref{fig:sys_figure}, further demonstrates the repeatable nature of these results.

A zoomed-in version of Fig.~\ref{fig:Triangle_5_0T_Loop1_Overlay} for just triangle pair 5 is shown in Fig.~\ref{fig:Triangle_5_0T_Loop1_Overlay}. An indication of the idealised bias triangle pairs is also shown, with an approximate fitting achieved by aligning with the back edge of the measured data. The base of the triangles corresponds to an alignment of the energy levels in the two quantum dots. The sides of the triangles then correspond to the level in one of the quantum dots aligning with the Fermi level in its adjacent reservoir~\cite{van_der_wiel_electron_2002}.

\section*{Conclusions}

We have demonstrated experimentally (and support through modelling and theory) the electrostatic control and existence of single and double quantum dot formation in the channel of a fully commercial FD-SOI device at 1~K. Biasing in any desired device state, as outlined in Tab.\,\ref{tab:device_states}, is achieved through a combination of common-mode voltage, back-gate voltage, and barrier gate voltages. Detuning of energy levels in the double quantum dot state is achieved through variations in the barrier gate voltages without a need for inter-barrier-gate plunger electrodes. There is a clear experimental demonstration of both single and double quantum dot charge transitions as predicted by our simulations. We demonstrate well-controlled single and double quantum dots with only barrier gate variations.  Accurate prediction and characterisation of quantum well formation in these commercial devices enables us to more precisely determine the quantum dot biasing conditions. This also allows for the potential future demonstration of qubit formation and charge sensing using a commercial device at 1~K.

\bibliography{references}

\newpage
\appendix
\counterwithin{figure}{section}
\section{QTCAD Model Calibration}\label{sup:model_calibration}

As discussed in the main text, the QTCAD model requires calibration, as it is very sensitive to a slight change of input parameters. One calibration method is to measure a DC current through the device and fit it with a simulated DC current. However, this is not feasable in cryogenic semiconductor devices, since the traditional Poisson equations are singular at $T\lessapprox 70$~K. QTCAD does not provide DC leakage current through the device, however, it can predict the biasing voltages of Coulomb blockade transitions and tunneling current using the WKB approximation and the Non-Equilibrium Greens Functions (NEGF) framework\cite{noauthor_qtcad_nodate}.

\begin{figure}[thb]
    \centering
    \includegraphics[width=0.9\textwidth]{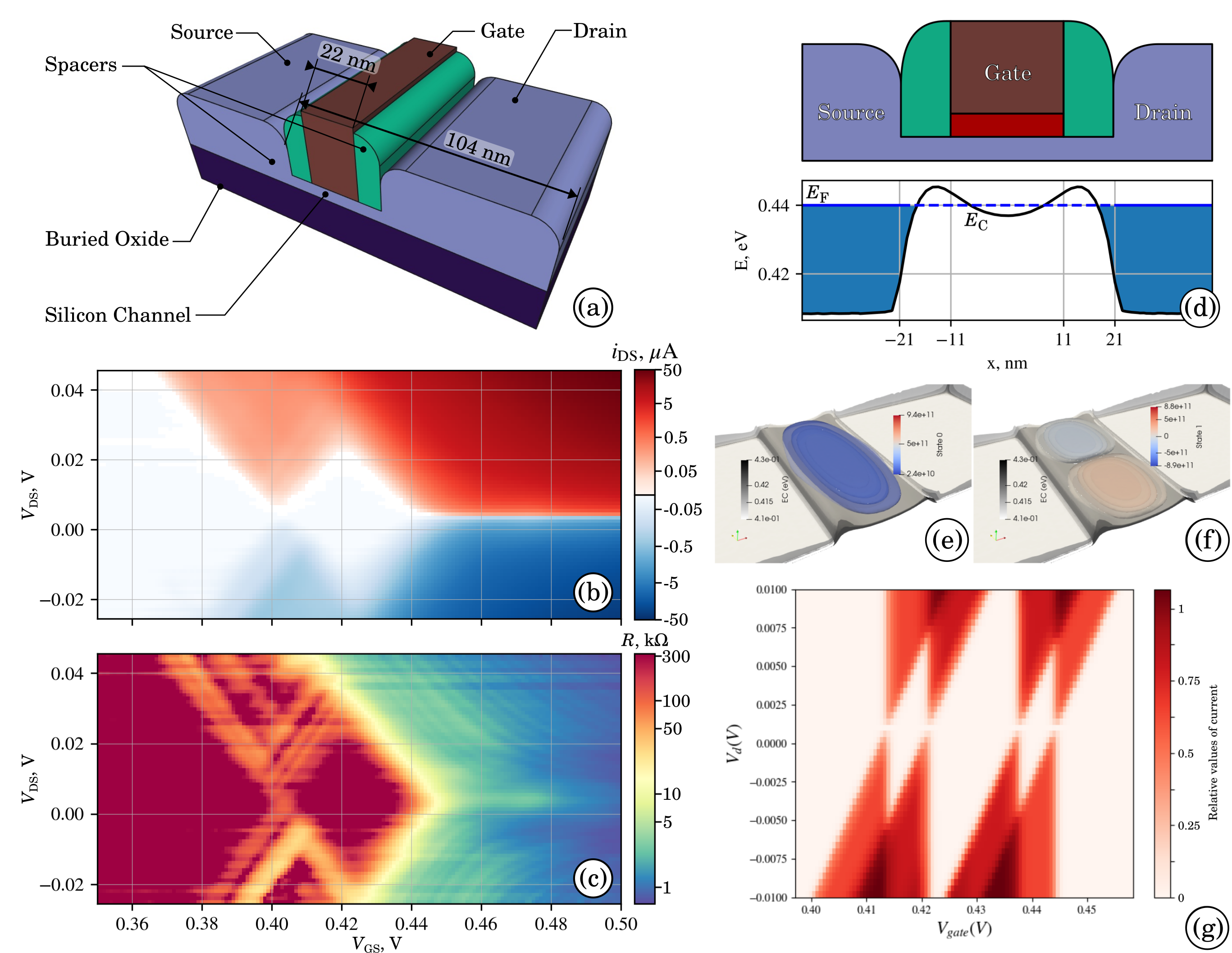}
    \caption{Experimental demonstration of the Coulomb blockade in the minimal feature size standard transistor. (a) The geometry of the transistor with a pitch of $104$~nm and gate width of $22$~nm. (b) The positive and negative drain-source DC current through the transistor (at $V_{\text{CM}} = -400~\text{mV}$). (c) Calculated resistance through the transistor. (d) The QTCAD simulation results for the minimal feature size FDX-22 transistor. 1D line cut of the conduction band aligned with the schematic view of the transistor. Blue areas correspond to the quantum leads. The black line is the conduction band edge. The blue dashed line is the Fermi level at the source and drain. (e) The wavefunction that corresponds to the ground state --- blue-to-red colours correspond to the wavefunction, and the grayscale colours --- equipotential surfaces. (f) The wavefunction that corresponds to the first excited state. (g) Coulomb diamonds were simulated using the master equation.}\label{fig:1gate_transistor_experiment}
\end{figure} 

The device considered in the main text is too large for calibration, given the size of its geometry and the number of input voltages, which increases the variability of results. Hence, experimental data from the $22~\text{nm}$ FD-SOI minimum size transistor (Fig.~\ref{fig:1gate_transistor_experiment}~(a)) was chosen. The DC measurements at $V_\mathrm{CM} = -400$~mV show a single clear Coulomb diamond, see Fig.~\ref{fig:1gate_transistor_experiment}~(b),(c). This implies that the device under test has a very shallow quantum well that can produce a single Coulomb blockade event before fully conducting. We then calibrated a model of this minimum size transistor to produce this same shallow well response, the calibrated parameters of which are then used in the full QDA model. We describe this calibration next.

The only possible configuration of the quantum well for the single-gate transistor is when the potential minimum is below the Fermi level and under the gate oxide, and potential barriers are higher than the Fermi level and forming below the spacers, see Fig.~\ref{fig:1gate_transistor_experiment}~(d). Therefore, as a first step, a sweep over the wide range of input parameters was applied:

\begin{itemize}
    \item Common mode voltage $-0.6~\mathrm{V} \leqslant V_\mathrm{CM} \leqslant -0.4~\mathrm{V}$, in steps of $20$~mV. 
    \item Gate-source voltage $0~\mathrm{V} \leqslant V_\mathrm{GS} \leqslant 0.65~\mathrm{V}$, in steps of $50$~mV. 
    \item Wide sweep of n-dopants concentration in the channel under the source and the drain. %
    \item Tuning of the gate and backgate workfunctions. 
\end{itemize}

The given parameters are chosen to be close to the expected ones. In total, this results in $9702$ simulations that are not feasible to analyse manually. Hence, the following automatic algorithm was used:
\begin{enumerate}
    \item Solve the Poisson equation with QTCAD at the given set of parameters. 
    \item Select three line cuts in the Silicon channel (at height $0.1$~nm, $3.0$~nm and $5.9$~nm from the top of the channel). 
    \item Define the maximal, $E_\mathrm{C}^\mathrm{max}$, and minimal, $E_\mathrm{C}^\mathrm{min}$, values of the conduction band edge (the minimal value is defined under the gate not including the source and drain regions). 
    \item Define the Fermi level, $E_\mathrm{F}$, under the source and drain (since the source/drain voltages are set by the shifting of the Fermi levels under the source and the drain). 
    \item The set of parameters is considered to be valid if $E_\mathrm{C}^\mathrm{min}<E_\mathrm{F}<E_\mathrm{C}^\mathrm{max}$.
\end{enumerate}
The given algorithm filtered $58$ valid sets of parameters of the $9702$ tested. An example of a valid 1D line cut of the conduction band is shown in Figure~\ref{fig:1gate_transistor_experiment}~(d). The conduction band edge that forms the dedicated quantum well makes it possible to calculate the single-electron wavefunctions, see Fig.~\ref{fig:1gate_transistor_experiment}~(e),(f). The depth of all analysed quantum wells is such that it can localize up to three quantum levels inside, all other wavefunctions will form outside of the barriers. This allows one to calculate the lever arm of the quantum dot and compare it with an experimental one ($\approx 0.8$eV/V, method outlined in\cite{venkatachalam_local_2012} and data shown in Figs.~\ref{fig:1gate_transistor_experiment}(b),(c)). The closest simulated lever arm was $\approx0.83$~eV/V, extracted using QTCAD. 

The final stage of the calibration was to match the position of the Coulomb peaks in simulations with one in experiments. QTCAD uses the master equation method to calculate the positions of the Coulomb peaks, and to reduce computational complexity it linearizes the single-electron energies in a quantum dot, which makes the shape of the Coulomb diamond straight, see Fig.~\ref{fig:1gate_transistor_experiment}~(g). All $58$ candidate sets of parameters were analysed and the closest parameters were used in the simulations of the minimum size transistor and the QDA in the main text.

\section{Operation Modes of The Quantum Dot Array}

\begin{figure}[htb]%
\centering
\includegraphics[width=0.99\textwidth]{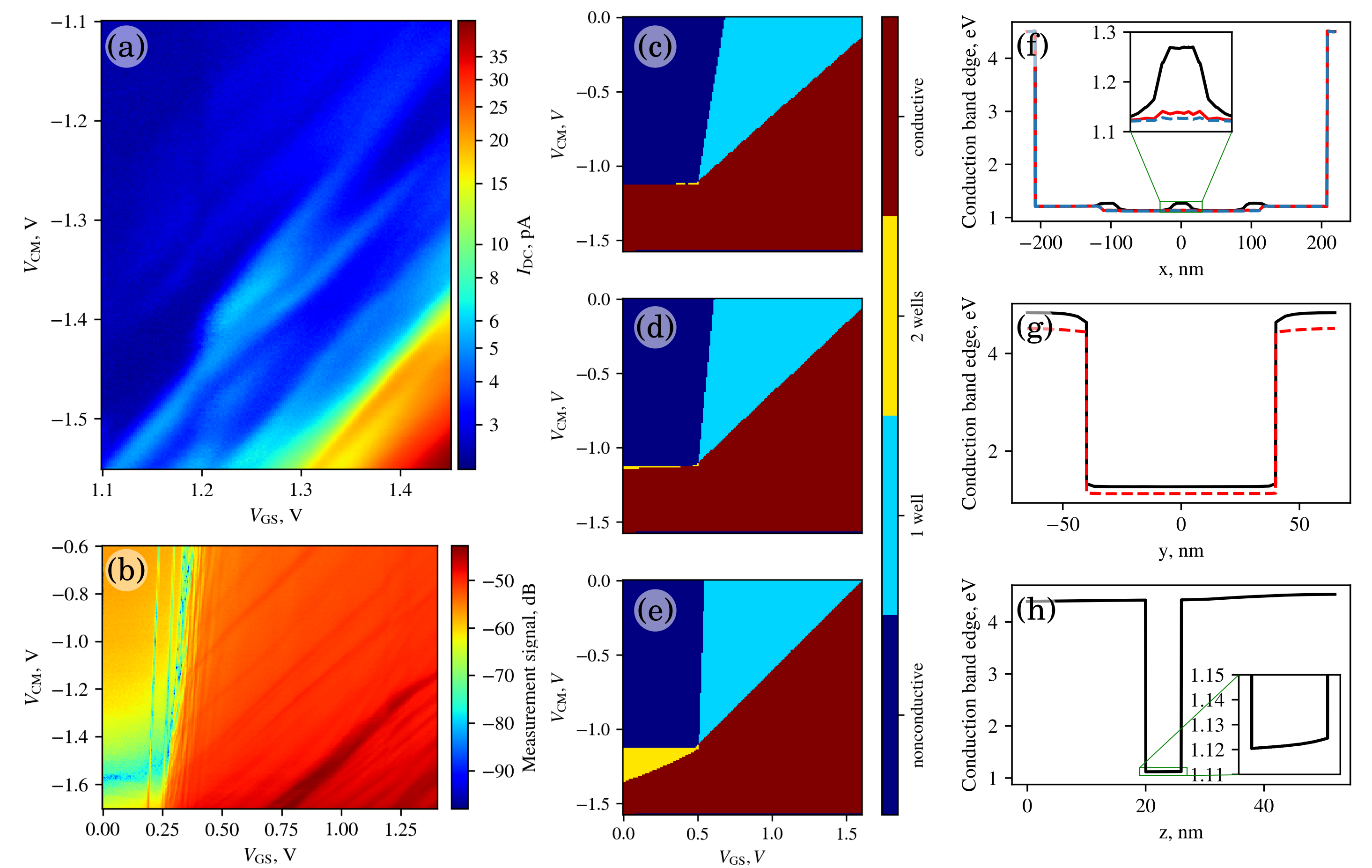}
\caption{The flat band experimental and simulation results on an older generation device with three barrier gates. (a) $3.6$~K DC measurements with $V_\mathrm{DS}=20$~mV. The same value of $V_{\text{GS}}$ was applied to all three barrier gates for this sweep (b) $3.6$~K RF reflectometry measurements with center frequency $f=236.255$~MHz, and $V_\mathrm{DS}=0$. (c) In the flat band QTCAD simulation results (built the same way as the one from the main text), the line cut is taken at the bottom of the silicon channel (close to the buried oxide). (d) A similar graph when the line cut is taken in the middle of the channel. (e) When the line cut is taken close to the top interface. (f) An example of the line cut graphs along the x-axis $V_\mathrm{GS}=200$~mV, $V_\mathrm{CM}=-1.25$~V. The solid black line is close to the interface, the solid red line is in the middle of the channel, and the blue dashed line is taken close to the buried oxide. The inset in the centre is a zoomed-in barrier. (g) The line cuts are taken along the y-axis at the same biasing condition. A solid black line is taken under the gate, the red dashed line is taken between the gates. (h) The z-cut was taken between the gates at the same biasing conditions. The inset is a zoomed-in section showing the bottom of the potential well. 
}\label{fig:flatband}
\end{figure}

One of the key experiments is the so-called flat band experiment. The idea of this experiment is to apply equal gate-source voltages to all gates with the negative common-mode voltage sweep. The experiments shown here were done on an older generation device with two techniques --- direct current measurements, see Fig.~\ref{fig:flatband}~(a), and RF-reflectometry measurements, see Fig.~\ref{fig:flatband}~(b). The DC measurements demonstrate the conduction-to-nonconduction transition, showing a diagonal pattern similar to the results shown in the main text flat band simulation. The RF-reflectometry doesn't have such a direct connection to the conductance of the structure but demonstrates the different well-configurations discussed in the main text. At low $V_\mathrm{GS}$ and high $V_\mathrm{CM}$, there is one type of pattern evident, but at higher $V_\mathrm{GS}$ and lower $V_\mathrm{CM}$ the pattern changes. This was our initial evidence of a transformation in the conduction band. 

The conduction band edge has a complicated 3D shape, therefore, it is not straightforward to analyse the data obtained from the QTCAD simulations. The confinement was classified at three different Z-level line cuts --- close to the buried oxide in Fig.~\ref{fig:flatband}~(c), in the geometrical centre of the silicon channel in Fig.~\ref{fig:flatband}~(d) and one that is close to the interface below the barrier gates in Fig.~\ref{fig:flatband}~(e). The biggest difference is the double-well (yellow) region, where wells are forming between the barrier gates in the channel. This happens because the potential barriers between wells vary strongly with the Z-direction, see Fig.~\ref{fig:flatband}~(f) showing X-direction line cuts at the three Z-cuts. Figure~\ref{fig:flatband}~(g) shows the Y-cut of the conduction bands under the barrier gates and between the barrier gates which shows strong confinement in this direction. Figure~\ref{fig:flatband}~(h) shows the Z-cut of the conduction band edge between the barrier gates. This shows strong confinement in the Z-direction, however, the bottom of a quantum well has a triangular shape, causing the wavefunction to be closer to the silicon oxide interface below the barrier gates. This may partly explain the high charge noise of this older generation device given the proximity to the upper oxide interface and charge trapping events. 

\section{Lever arm of the device in different operation modes.}

\begin{figure}[htb]%
\centering
\includegraphics[width=0.49\textwidth]{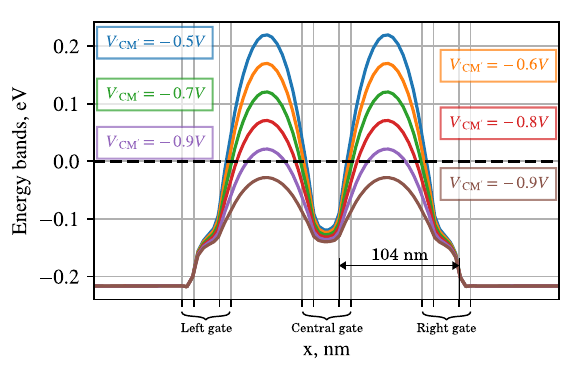}
\caption{The effect of the common-mode voltage on the single-well configuration. }\label{fig:vcm_single_gate}
\end{figure}

\begin{figure}[htb]%
\centering
\includegraphics[width=0.49\textwidth]{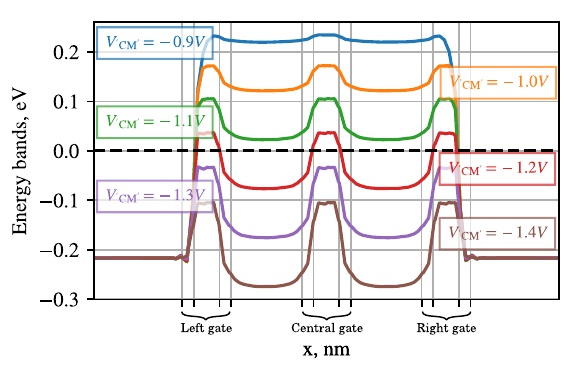}
\caption{The effect of the common-mode voltage on the double-well configuration. }\label{fig:vcm_double_gate}
\end{figure}

\begin{figure}[htb]%
\centering
\includegraphics[width=0.49\textwidth]{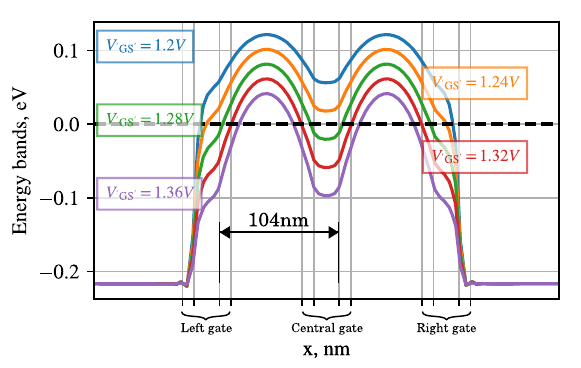}
\caption{The effect of all-gates voltages shift $V_\mathrm{GS}=V_\mathrm{GS\,left}=V_\mathrm{GS\,central}=V_\mathrm{GS\,right}$ in a single-well configuration. }\label{fig:vgs_single_gate}
\end{figure}

\begin{figure}[htb]%
\centering
\includegraphics[width=0.49\textwidth]{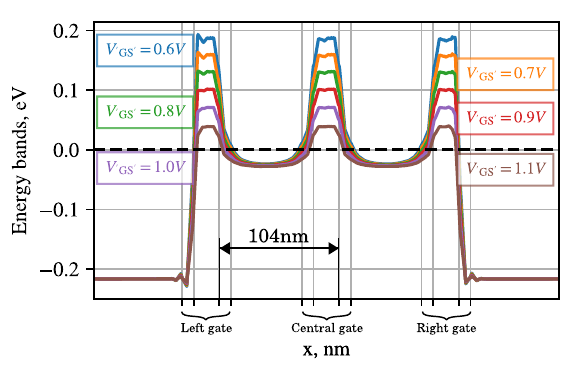}
\caption{ The effect of all-gates voltages shift $V_\mathrm{GS}=V_\mathrm{GS\,left}=V_\mathrm{GS\,central}=V_\mathrm{GS\,right}$ in a double-well configuration.}\label{fig:vgs_double_gate}
\end{figure}

It is very important to understand how the voltage applied to the gates acts on the confinement at the different modes (1-well and 2-wells). The standard approach in this case is to calculate the lever arm of the gates, however, as it was explained in the main text, the biasing was done by applying the combinations of the voltages. Therefore, to have an orientation in the effect of the applied biasing we build the set of graphs (Fig.~\ref{fig:vcm_single_gate},~\ref{fig:vcm_double_gate},~\ref{fig:vgs_single_gate},~\ref{fig:vgs_double_gate}). 

The common mode voltage effect is visible in Fig.~\ref{fig:vcm_single_gate}. Note that all graphs are normalized to the Fermi-energy shift to compare the similar conduction bands. It is easy to see that the barriers that are forming between gates are more affected by the change of the $V_\mathrm{CM}$, while the energy of the well changes not that much. 

\section{Measurement Setup}

\begin{figure*}[h]
    \centering
    \includegraphics[width=0.99\textwidth]{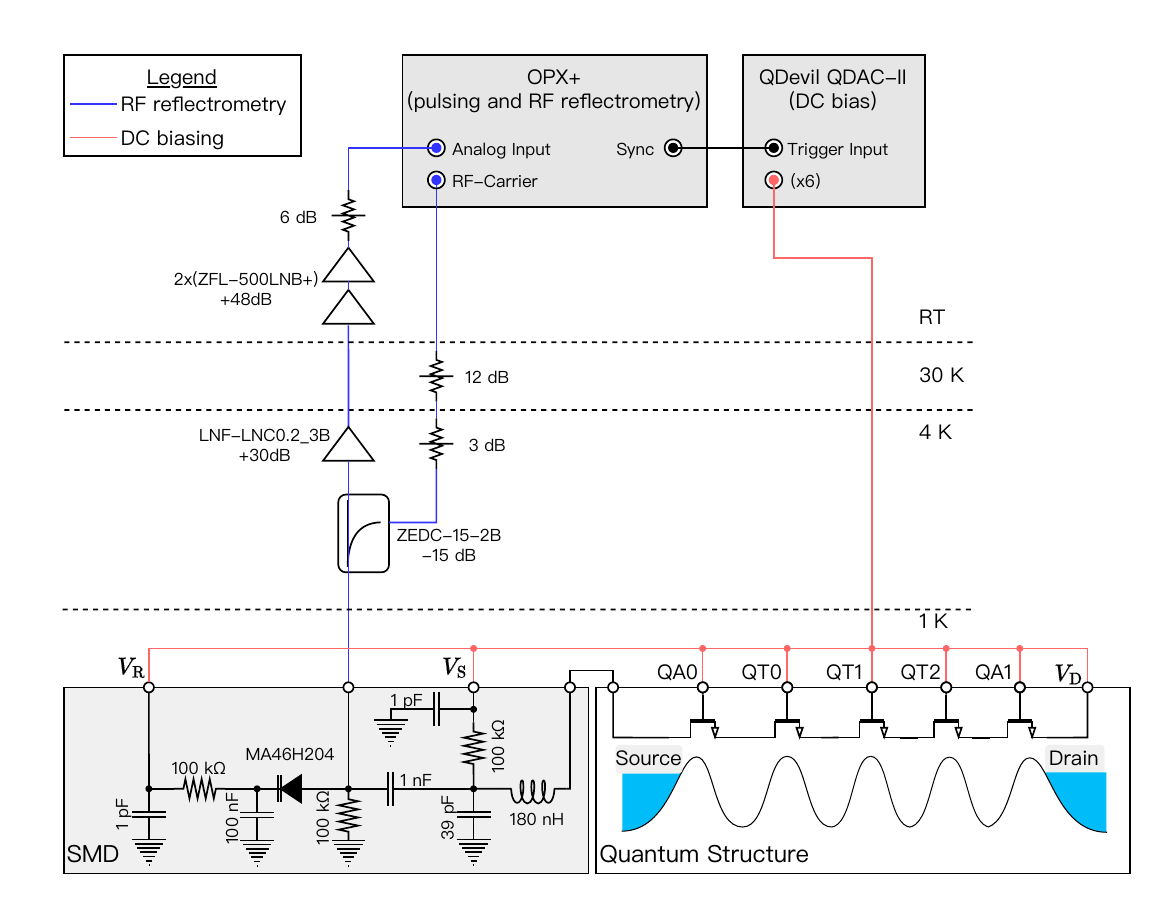}
    \caption{Experimental setup. The cryostat has a base temperature of $1.0$~K. A QDevil QDAC-II was utilized to generate the DC signals at low frequencies. A Quantum Machines OPX+ is used for RF reflectometry. }
    \label{fig:exp_circuit}
\end{figure*}

The RF carrier frequency used is approximately 264MHz and the power of the RF signal is optimized to -50dBm for clearest reflectometry signals. The signal is attenuated at both the $40$~K and the $4$~K stages of the refrigerator to reduce thermal noise and then also further attenuated by the directional coupler. The signal power incident on the PCB sample board is approximately -80dBm. %
The RF carrier is reflected by the tank circuit on the PCB sample board and amplified by a cryogenic amplifier at 4 K after the directional coupler. The RF carrier is then amplified again at room temperature, and digitized at 1 GS/s.
DC- and low-frequency voltages are generated using a QDevil QDAC-II\cite{noauthor_quantum_nodate} high-precision low-noise digital-to-analogue converter and transmitted through an 84-pin flex cable from room temperature to the 1K stage. The QDAC-II is triggered by the OPX+, and the DC voltages are loaded to the QDAC-II in advance and swept by trigger signals from the OPX+. This allows for synchronization between DC voltage sweeps and RF reflectometry readout. The experimental setup employed in this study is shown in Fig.~\ref{fig:exp_circuit}.

\section*{Acknowledgements}

This work was done under the Disruptive Technologies Innovation Fund QCoIr EI Project, Ireland, 166669/RR.

\end{document}